\documentclass[final,5p,twocolumn]{elsarticle}
\usepackage{amssymb}
\usepackage{amsthm}
\usepackage{todonotes}
\usepackage{tikz}
\usetikzlibrary{automata,positioning}
\usepackage{mathtools}
\usepackage{MnSymbol}

\usepackage{booktabs}       
\usepackage{amsfonts}       
\usepackage{nicefrac}       
\usepackage{microtype}      

\usepackage{graphicx}
\usepackage{wrapfig}
\usepackage{subfig}

\usepackage{amsmath}

\usepackage{textcomp}

\usepackage{algorithm}
\usepackage{algorithmic}

\newcommand{\RNum}[1]{\uppercase\expandafter{\romannumeral #1\relax}}

\usepackage{mathtools}

\newtheorem{assumption}{Assumption}

\usepackage{natbib}

\newcommand{\argmaxF}{\mathop{\mathrm{argmax}}\limits}   
\newcommand{\argminF}{\mathop{\mathrm{argmin}}\limits}   

\usepackage{enumitem}
\setitemize{nolistsep}
\setenumerate{nolistsep}

\journal{}

\newtheorem{rem}{Remark}

\newtheorem{ex}{Example}

\numberwithin{equation}{section}

\newcommand{\abs}[1]{\left\lvert{#1}\right\rvert}

\renewcommand{\geq}{\geqslant}

\renewcommand{\leq}{\leqslant}

\newcommand{\R}{\mathbb{R}}

\allowdisplaybreaks

\DeclareMathOperator{\sbjto}{subject\:to}

\begin{document}

\begin{frontmatter}

\title{A reinforcement learning approach to hybrid control design}

\author[A1]{Meet Gandhi}
\author[A2]{Atreyee Kundu}
\author[A3]{Shalabh Bhatnagar}

\address[A1]{Department of Computer Science and Automation, Indian Institute of Science Bangalore, India, Email: meetgandhi@iisc.ac.in}
\address[A2]{Department of Electrical Engineering, Indian Institute of Science Bangalore, India, Email: atreyeek@iisc.ac.in}
\address[A3]{Department of Computer Science and Automation, Indian Institute of Science Bangalore, India, Email: shalabh@iisc.ac.in}



\begin{abstract}
    {In this paper we design hybrid control policies for hybrid systems whose mathematical models are unknown. Our contributions are threefold. First, we propose a framework for modelling the hybrid control design problem as a single Markov Decision Process (MDP). This result facilitates the application of off-the-shelf algorithms from Reinforcement Learning (RL) literature towards designing optimal control policies. Second, we model a set of benchmark examples of hybrid control design problem in the proposed MDP framework. Third, we adapt the recently proposed Proximal Policy Optimisation (PPO) algorithm for the hybrid action space and apply it to the above set of problems. It is observed that in each case the algorithm converges and finds the optimal policy.}
\end{abstract}

    \begin{keyword}
        Hybrid Control, Markov Decision Process, Reinforcement Learning, Proximal Policy Optimisation, Neural Networks
    \end{keyword}
\end{frontmatter}
\section{Introduction}
\label{s:intro}
\subsection{Motivation}
\label{ss:motive}
    {A \emph{hybrid dynamical system} involves interaction between continuous (time-driven) and discrete (event-driven) dynamics. A large class of real-world systems including air traffic management, chemical process control, communication networks, embedded control, engine control, and robotics can be modelled as hybrid systems. In this paper we focus on the design of hybrid control policies for hybrid systems. Such control policies comprise of both continuous and discrete signals, which take values in a continuum (such as the set of real numbers \(\R\)) and a finite set (such as \(\{a, b, c\}\)), respectively. The reader is referred to \cite{Johansson_book,LygerosChapter} for detailed discussions on hybrid systems and control.}
    
    {The problem of designing hybrid control policies that ensure good qualitative and quantitative properties of systems has attracted considerable research attention in the past two decades. This includes stabilization and optimal control of power electronic devices \cite{Thomas2015}, mobile robots \cite{Bak2003}, autonomous vehicles \cite{Kapania2019}, etc. A common assumption in the existing body of works is the availability of accurate mathematical model of the system that is to be controlled. However, system identification of hybrid systems, even in the simplest form of switched linear systems or piecewise affine systems, is known to be a NP-hard problem, see e.g., \cite{Lauer2016} for a detailed discussion. This fact motivates the current paper.} 
    
    {We consider hybrid systems whose underlying mathematical models are not available and design hybrid control policies that optimize certain performance criteria of the systems.} In the sequel we will occasionally abbreviate a hybrid control policy as a control policy whenever there is no risk of confusion. We will also call a control policy that optimizes the desired performance criteria as an optimal control policy.
\subsection{Prior works}
\label{ss:lit_survey}
    {A natural tool for model-free design of optimal control policies is Reinforcement Learning (RL) algorithms. Indeed, the close connection between RL and optimal control are well-known, see e.g., \cite[Chapter 3]{SuttonBarto_book}, \cite{Bert1} for details. 
    
    In \cite{Dong2003} the authors apply learning automata towards designing optimal control policies for hybrid systems. This technique was later extended to a blend of learning automata and RL algorithms in \cite{JXIa}. {In \cite{Li} a hybrid control policy that minimizes a certain quadratic state-input cost for switched linear systems is designed by employing a Q-learning based technique.} {The recent work} {{\color{red}}\cite{Deepmind} proposes an RL algorithm with hybrid (continuous and discrete) action space that designs optimal control policies for hybrid systems with controlled and/or autonomous switching. A value iteration based RL technique has been employed to design optimal control policies for stochastic hybrid systems with jumps in \cite{MDP2}. The proposed {method}, however, requires complete knowledge of the system model. In this paper we target an RL-based design of optimal control policies for a larger class of hybrid systems.}
\subsection{Our contributions}
\label{ss:contri}
     We consider two types of hybrid systems, commonly classified as time-driven systems that include event-driven dynamics and event-driven systems that include time-driven dynamics, see e.g., \cite{Branicky1, Cassandras1} for details. The former type refers to systems where event-driven dynamics arises as substructures of time-driven dynamics, while the latter type refers to systems where time-driven dynamics arises as substructures of event-driven dynamics. In the sequel we will call these types of hybrid systems as Type-I and Type-II hybrid systems and the corresponding problems of designing hybrid control policies as Type-I and Type-II hybrid control design problems, respectively. We use two running examples to illustrate the complexities of the {above classes of} hybrid control design problems.
     \begin{ex}{ (Type-I hybrid control design problem: Four-gear car)}
    \label{ex:four-gear_car}
    \rm{
        {We consider the setting of a four-gear car from \cite{LygerosChapter}}. Let $x_1$, $x_2$ and $q$ denote the longitudinal position, velocity and the gear position of the car respectively. The dynamics of the system is driven by
        \begin{equation}
            \begin{bmatrix}
                \dot{x}_{1,t}       \\
                \dot{x}_{2,t}
            \end{bmatrix}
            =
            \begin{bmatrix}
                x_{2,t}     \\
                f_{q}(x_{2,t})u_t
            \end{bmatrix},
        \label{Eq_CG_Model1}
        \end{equation}
        where the function $f_{q}(x_2)$ {gives the acceleration generated at a particular gear and velocity pair, \((q,x_2)\) and its behaviour for different gears is shown in figure \ref{CG_Velocity_Acceleration1}}.
        \begin{figure}[htbp]
        \centering
            \includegraphics[scale=0.5]{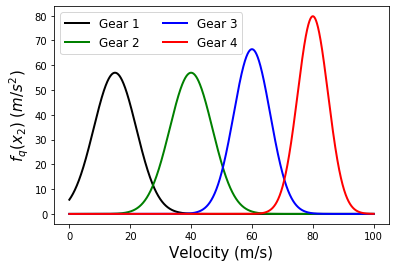}
            \caption{Acceleration as a function of velocity for different gears.}
            \label{CG_Velocity_Acceleration1}
        \end{figure}

        At all velocities, appropriate gear selection is essential. Otherwise, little acceleration will be generated and the car's velocity will not increase. At every {instant of time}, the controller has two decisions to take: the throttle position $u \in [0,1]$, which is a continuous control {action} (to decide what percentage of acceleration to use), and $v \in \{-1,0,1\}$, which is a discrete control {action} (to decide whether to decrease the gear by 1, ride in the same gear or increase the gear by 1). Decreasing the gear when the car is in gear $1$ and increasing the gear when it is in gear $4$ is not permitted. When the car reaches the velocity of 90 m/s, a sudden brake is automatically applied, which resets its velocity to 20 m/s. Starting the stationary car in the first gear, the goal of the controller is to cover maximum distance in {a stipulated time, say} $10$ {seconds} by controlling the throttle and the gear box.

        {Notice that} we have a collection of four dynamical systems, one for each gear. At time $t$, the continuous state, $(x_{1,t}, x_{2,t})$, represents the position and velocity of the car, which is governed by \eqref{Eq_CG_Model1}, {while} the discrete state $q_t$ represents the gear of the car. We make two observations: (i) The system \emph{switches} from one dynamical system to another when a gear is changed. After the switch, the continuous state starts evolving according to a \emph{new} dynamical system. (ii) A sudden \emph{impulse} is observed (velocity reset to 20 m/s) when the continuous state reaches a certain set (velocity reaches 90 m/s). \qed
    }
    \end{ex}
    \begin{ex}{(Type-II hybrid control design problem: Steel annealing process)}
    \label{ex:steel-annealing_process}
    \rm{
        {We consider the setting of a steel annealing process from \cite{SteelAnnealing}.} Several strips pass through a heating furnace, that changes their temperature.  Each strip (referred to as a job) has two states: a physical state (temperature), and a temporal state (arrival and departure time). The temperature, $x(t)$, starting from 30\textcelsius{}, changes according to {the following dynamics}:
        \begin{equation}
            \dot{x}(t) = -K_1u + K_2(F^4 - x(t)^4).
            \label{Eq_SA_non_linear_heat_transfer1}
        \end{equation}
        The rate of change of temperature depends on the strip velocity, $u$ (remains constant during the process), and the furnace reference temperature, $F$ (which is 600\textcelsius{}). Here, $K_1 = 1.14$ and $K_2 = 7.82\times10^{12}$ are prescribed constants. The complex non-linear heat transfer equation \eqref{Eq_SA_non_linear_heat_transfer1} is approximated by
         \begin{equation}
            \dot{x}(t) = \frac{1}{f(u)}(F - x(t)),
            \label{Eq_SA_approx_heat_transfer1}
        \end{equation}
        where $f(u) = 5.4u + 29$.

        Consider $N$ strips, where the $i^{\text{th}}$ strip has length $h_i$ and arrives at time $a_i$. The \emph{temporal state}, $z_i$, of the $i^{\text{th}}$ strip is denoted by the tuple \((z_{1,i}, z_{2,i})\), where $z_{1,i}$ and $z_{2,i}$ represent the arrival and the departure times of the job $i$ respectively. Suppose {that} the controller sets the velocity of the $i^{\text{th}}$ strip to $u_i$. The temporal states $z_i$ are computed as
        \begin{equation}
            \begin{aligned}
                z_{1,i} = \max(a_i, z_{1,i-1}) + \frac{h_i}{u_i}; \\  z_{2,i} = z_{1,i} + \frac{L}{u_i},
            \end{aligned}
            \label{Eq_SA_Temporal_State1}
        \end{equation}
        where $z_{1,0} = -\infty$ and $L = 500$m.

        {We have that} $u_{\min} \leq u \leq u_{\max}$. {In view of} \cite[Lemma 1]{SteelAnnealing}, if $\beta = 1/f(u_i)$, then $\theta(u_i)$ (resp. $\phi(z_i)$), defined {as}
         \begin{equation}
            \begin{aligned}
                \phi(z_i)  &=  -(z_{2,i})^2, \\
                \theta(u_i) &=  -\abs{F - x(L/u_i)}^2 - \beta \int_0^{L/u_i}(F - x(t))^2 dt,
            \end{aligned}
            \label{Eq_SA_Costs1}
        \end{equation}
        is a decreasing (resp. increasing) function of $u_i$. The agent here needs to strike a balance between two competing objectives: (i) {reducing} temperature error with respect to the furnace reference temperature $F$, and (ii) {delivering} the completed jobs in timely fashion. Thus, the optimal control problem is given {as follows}:
        \begin{equation}
            \begin{aligned}
                \max_{u_1, .. ,u_N} \sum_{i=1}^{N} &\:\: [{ \phi(z_{i}) + \theta(u_i)}] \\
                \sbjto &\:\:
                (\ref{Eq_SA_non_linear_heat_transfer1}), (\ref{Eq_SA_Temporal_State1}), (\ref{Eq_SA_Costs1}).
            \end{aligned}
            \label{Eq_SA_Optimisation_Equation1}
        \end{equation}\qed
        }
        \end{ex}

        {We observe the following differences between the two settings described above:} (i) The first {example} deals with only one car, {while} the second {example} involves several strips. (ii) In the first {example} the controller has to manage the throttle and the gear box during the entire $10$s period, {while} in second {example} the controller observes the arrival times of strips and decides their (time-independent) velocities in one go. (iii) In the first {example} the car has only the physical state (represented by the position, velocity and the gear), {while} in the second {example} each strip has a physical state (temperature) as well as a temporal state (arrival and departure times).
        
        {Our key contributions in this paper are the following}:
        \begin{itemize}[label = \(\circ\), leftmargin = *]
            \item First, we model both Type-\RNum{1} and Type-\RNum{2} hybrid control design problems in a MDP framework. {This result facilitates the application of off-the-shelf algorithms from RL literature towards designing optimal hybrid control policies for hybrid dynamical systems whose mathematical models are not available.}
            \item Second, {we model five benchmark hybrid control design problems in the proposed MDP framework.} 
            \item Third, we {adapt} the Proximal Policy Optimisation (PPO) algorithm for hybrid action space and apply it on {the above set of} problems. {It is observed in each case that the algorithm converges and finds the optimal hybrid control policy.}
        \end{itemize}
        We emphasize that our {design of hybrid control policies} does not require an explicit knowledge of a {mathematical model of the system that is to be controlled, and} hence, is an instance of a model-free approach for {the design of optimal hybrid control policies}. Our results differ from the existing techniques in terms of the classes of hybrid systems under consideration and the algorithm employed for the design. In fact, to the best of our knowledge, this is the first instance in the literature where {an actor-critic algorithm (PPO in this case)} is applied to address hybrid control design for a vast class of hybrid systems. We employ the OpenAI Gym API platform for the implementation task \citep{OpenAI_Gym}.\footnote{Our codes along with instructions for executing them are available as supplementary material with this manuscript.}
\subsection{Paper organization}
\label{ss:paper_org}
        The {remainder} of the paper is organised as follows: in \S\ref{s:prelims}, we {catalog a set of descriptions and notions from RL literature. Our results appear in \S\ref{s:mainres}. We also describe various features of our results in this section. We conclude in \S\ref{s:concln} with a brief discussion of future research directions.}
\section{Preliminaries}
\label{s:prelims}
\subsection{Reinforcement Learning, Markov Decision Processes and Proximal Policy Optimisation algorithm}
\label{ss:RL_prelims}
    RL \cite{NDP_book, SuttonBarto_book} uses the framework of MDP to define interactions between a learning agent and its environment. The goal of the agent is to maximise a certain utility or a long-term reward. An MDP consists of the following components: a state space $\mathbb{S}$, an action space $\mathbb{A}$, a model $p(s_{t+1}|s_t, a_t)$, which specifies the probability of transition from a state $s_t$ to another state $s_{t+1}$ under the action $a_t$, a reward function $r(s_t, a_t, s_{t+1}) \in \mathcal{R}$ and the discount factor, $\gamma \in [0,1)$, which ensures that the total return is bounded. The agent samples actions from a state-conditioned distribution function $\pi(a|s)$. The goal of the agent is to find a certain policy $\pi$ to maximise the expected return or the value function $J_{\pi}(s) \coloneqq \mathbb{E}_{\pi} [\sum\limits_{t=0}^\infty \gamma^tr(s_t, a_t)|s_0=s]$ when starting from any given state $s$.

    RL algorithms can be broadly classified into three categories: value-based algorithms, policy-based algorithms and actor-critic algorithms. For the value-based algorithms, for some policy $\pi$, we define the action value function, $Q^{\pi}(s,a)$, as the expected discounted reward when choosing the action $a$, in the state $s$, and then acting subsequently according to the policy $\pi$. Thus, $Q^{\pi}(s,a) \coloneqq \mathbb{E}_{\pi}[\sum\limits_{t=0}^\infty \gamma^tr(s_t, a_t)| s_0=s; a_0=a]$. Our aim is to find $\pi^{*}$, such that $J_{\pi^{\text{*}}}(s)$ is maximum for any $s$. While the agent is in state $s$, it computes $Q^{\pi\text{*}}(s,a)$ for all allowed actions in the state $s$, and chooses the action with the maximum action value. Thus, $J_{\pi^{*}}(s) = \max\limits_a Q^{\pi^{*}}(s,a)$. However, we do not explicitly compute $\pi^{\text{*}}$ in value-based algorithms. In the policy-based algorithms, we directly try to find $\pi^{*}(a|s)$ and do not compute the action-value function $Q^{\pi}(s,a)$ for a given policy $\pi$. We restrict the policy to a certain class, parameterised by theta: $\pi_{\theta}(a|s)$. Our modified problem restricts the search in the domain of $\theta$ and tries to find a certain $\theta$ that maximises $J(\pi_{\theta})$. Actor-critic algorithms combine features of both the above methods. They involve both: an explicit policy (actor) update and an action value function (critic) update. The PPO algorithm is a recent actor-critic algorithm \cite{PPO}; {we provide its pseudo-code} in Algorithm \ref{PPO}.

    Most of the effective RL algorithms have their roots in Deep Learning \cite{policy_gradient, trpo, PPO}.
    Several earlier algorithms that use linear function approximators provided the guarantees of convergence \cite{TD0Convregence1, TD0Convregence2}. However, the linear function approximators are limited by their capability. This is precisely where the recent deep learning framework is helpful. It allows us access to highly non-linear function approximators. But it has two major problems: (a) It is hard to understand the hidden features that a neural network learns. (b) We lose the convergence guarantees that linear function approximators provide \cite{Baird, TD0Convregence1}. Hence, the reflection of the state-of-the-art in Reinforcement Learning (RL) is: \emph{There are no methods that are guaranteed to work for all or even most problems, but there are enough methods to try on a given problem with a reasonable chance of success in the end \cite{Bert1}.} PPO is one such algorithm, which uses deep neural networks as actor and critic architectures. Note that convergence of the PPO algorithm to an optimal policy is not guaranteed mathematically. However, {for all our benchmark examples, it is observed that the algorithm converges and finds an optimal hybrid control policy}.

    \begin{algorithm*}
    \caption{Pseudo-code for PPO algorithm}\label{PPO}
    \begin{algorithmic}[1]
        \STATE Input: initial policy parameters $\theta_0$, initial value function parameters $\phi_0$.
        \STATE \textbf{for} k = 0, 1, 2,\ldots \textbf{training epochs}:
        \STATE\hspace{\algorithmicindent} Collect set of trajectories $\mathbb{D}_k$ = $\{\tau_i\}$ by running policy $\pi_k$ = $\pi(\theta_k)$
        \STATE\hspace{\algorithmicindent} Compute Rewards-to-go $\hat{R}_t$.
        \STATE\hspace{\algorithmicindent} Compute Advantage Estimates $\hat{A}_t$ using any method of advantage estimation based on $V_{\phi_k}$
        \STATE\hspace{\algorithmicindent} Update the policy by maximising the PPO-Clipped Objective via some gradient descent algorithm:
        \STATE\hspace{\algorithmicindent}\hspace{\algorithmicindent}\hspace{\algorithmicindent} $\theta_{k+1} = \argmaxF_{\theta} \frac{1}{|\mathbb{D}_k|T}\sum\limits_{\tau \in \mathbb{D}_k}\sum\limits_{t=0}^{T}\min(\frac{\pi_{\theta}(a_t|s_t)}{\pi_{\theta_k}(a_t|s_t)}\mathbb{A}^{\pi_{\theta_k}}(s_t, a_t), g(\epsilon, \mathbb{A}^{\pi_{\theta_k}}(s_t, a_t)))$
        \STATE\hspace{\algorithmicindent} Update the critic network by maximising the MSE via some gradient descent algorithm:
        \STATE\hspace{\algorithmicindent}\hspace{\algorithmicindent}\hspace{\algorithmicindent}\hspace{\algorithmicindent}\hspace{\algorithmicindent}\hspace{\algorithmicindent}\hspace{\algorithmicindent}\hspace{\algorithmicindent} $\phi_{k+1} =
                \argminF_{\theta} \frac{1}{|\mathbb{D}_k|T}\sum\limits_{\tau \in \mathbb{D}_k}\sum\limits_{t=0}^{T}(V_{\phi}(s_t) - \hat{R}_t)^2$
        \STATE \textbf{End for}
    \end{algorithmic}
    \end{algorithm*}
\subsection{Our implementation of PPO algorithm}
\label{ss:PPO_implement}
    We adapt the spinning up implementation of PPO for the hybrid action space \cite{SpinningUp2018}.
    We assume that all actions are independent of each other {and} that all discrete actions follow the Boltzmann distribution and all continuous actions follow the Normal distribution (all continuous actions have the same variance). To construct a policy for the hybrid action space, we have a \textit{single} {neural} network that outputs {the} means for the continuous actions and category weights for discrete actions. For Example 1, with one discrete action, $v$ (that can take $3$ different values), and one continuous action, $u$, the actor {neural} network has four output nodes: the first three nodes output the weights corresponding to different categories of $v$ (using which the Boltzmann distribution is constructed) and the last node outputs the mean of $u$ (using which a Normal distribution is constructed). The critic {neural} network has the same architecture as the actor {neural} network (two hidden layers, each containing $64$ nodes with \textit{Tanh} activation), except that it has a single output node which gives an estimation of the state value of a certain state. {Note that we have used the same actor and critic network architectures (two hidden layers, each layer contains 64 nodes, all nodes have \textit{Tanh} activation) for all problems}. {Note that we have used the same number of nodes in each hidden layer as well as the same activation function for the actor as well as the critic neural network for all our benchmark problems.
    We are now in a position to present our results.
\section{Results}
\label{s:mainres}
    We describe our results for Type-I and Type-II hybrid control design problems separately.
\subsection{Type-\RNum{1} hybrid control design problem}
\label{ss:mainres1}
    {Recall that a hybrid dynamical system involves an interaction between continuous (time-driven) and discrete (event-driven) dynamics. The time-driven dynamics involves a continuous state process (denoted by $x$) that evolves according to some differential equations, and the event-driven dynamics involves a discrete state process (denoted by $q$) that is driven by some discrete event system. A natural analytical complexity for these systems arises due to a} ``jump'' from one state to another. Two types of jumps are possible: (i) switching and (ii) impulse. After a switch, the dynamics $q$, and thus the differential equation changes abruptly, and from then on, the system starts evolving according to a new differential equation, whereas an impulse changes the continuous state $x$ abruptly. {The} jumps can {as well} occur in two modes: controlled and autonomous. {A \emph{controlled} jump is caused by an external input/action, while an autonomous jump is governed by internal mechanism of the system}. {Consequently}, any modelling framework needs to cover the {following} four phenomena: (i) controlled switching, (ii) controlled impulse, (iii) autonomous switching and, (iv) autonomous impulse. Example \ref{ex:four-gear_car} is an instance of {controlled switching with autonomous impulse}. After \textit{the controller decides} to change the gear, velocity $x_2$ evolves according to a different dynamics $f_q(x_2)$, resulting in controlled switching. The car's velocity \textit{automatically} resets to 20 m/s after reaching 90 m/s, implying an autonomous impulse.

    The work \cite{Branicky1} proposes a framework to model the classes of problems described above as controlled general hybrid dynamical systems (CGHDS) described by the subtuple:
    $H_c = [\mathbb{Q}, \Sigma, \mathbb{B}, \mathbb{G}, \mathbb{V}, \mathbb{C}, \mathbb{D}]$. Here, $\mathbb{Q}$ is a set of discrete states where $q \in \mathbb{Q}$ represents the index of the dynamics.
    $\Sigma = \{\Sigma_q\}_{q \in \mathbb{Q}}$ is a collection of controlled dynamical systems, where each $\Sigma_q = (x_q, \dot{x}_q, u_q)$ represents a continuous state space, a continuous dynamics and a set of continuous control. {We have} $\dot{x}_q: x_q \times u_q \rightarrow x_q$. Further, $\mathbb{V}$ is a set of discrete controls. $\mathbb{B} = \{B_q\}_{q \in \mathbb{Q}}$ and $\mathbb{C} = \{C_q\}_{q \in \mathbb{Q}}$ are collections of autonomous jump sets and controlled jump sets respectively, where $B_q, C_q \subset x_q$ for all $q \in \mathbb{Q}$. Also, $\mathbb{D} = \{D_q\}_{q \in \mathbb{Q}}$ is a collection of destination sets with $D_q \subset \mathbb{S}$ and $\mathbb{S} = \bigcup\limits_{q \in \mathbb{Q}} x_q \times \{q\}$. $\mathbb{G} = \{G_q\}_{q \in \mathbb{Q}}$ is a collection of autonomous jump transition maps, where $G_q: B_q \times \mathbb{Q} \rightarrow \mathbb{S}$. The total reward is given by
    \begin{align}
            R &= \int \limits_{0}^{\infty} e^{-at}K(x(t), q(t), u(t))\, dt\nonumber\\
            &+\sum \limits_i e^{-a\sigma_i}R_{b,i}(x(\sigma_i), q(\sigma_i), v_i)\nonumber\\
            &+\sum \limits_j e^{-a\zeta_j}R_{c,j}(x(\zeta_j^{-}), q(\zeta_j^{-}), x(\zeta_j^{+}), q(\zeta_j^{+})),
        \label{Eq_HC_Total_Reward}
    \end{align}
    where $a > 0$ is the discount factor, $\{\sigma_i\}$ and $\{\zeta_j\}$ represent the autonomous and controlled jump times respectively. $x(\zeta_j^{-})$, $q(\zeta_j^{-})$ and $x(\zeta_j^{+})$, $q(\zeta_j^{+})$ respectively represent the pre and post states during the $j^{\text{th}}$ controlled jump. $K(\cdot, \cdot, \cdot)$ denotes the operating reward, $R_{b}(\cdot, \cdot, \cdot)$ the autonomous jump reward, and $R_{c}(\cdot, \cdot, \cdot, \cdot)$ is the controlled jump reward.

    In Example \ref{ex:four-gear_car} we have $\mathbb{Q} = \{1,2,3,4\}$, $x_q = \mathbb{R}^2$, $\dot{x}_q$ is given by \eqref{Eq_CG_Model1}, $u_q = \{u\}$, $\mathbb{V} = \{v\} $.
$B_q = \mathbb{R} \times \{90\}$ and $\mathbb{C}_q = \mathbb{R}^2$ for all $q \in \mathbb{Q}$. $\mathbb{D}_{q(t)} = (x_1(t), x_2(t)) \times (\{q(t)-1, q(t), q(t)+1\} \cap \{1, 2,3,4\}) $. $G_q(x_1, 90, q) = (x_1, 20, q)$ for all $q \in \mathbb{Q}$. $K(x(t), q(t), u(t)) = x_2(t)$ and both $R_b$ and $R_c$ are zero.
    \subsubsection{MDP Framework}
        Under the assumptions mentioned below, we propose an MDP framework for {Type-I hybrid control design} problems.
        \begin{assumption}
            All dynamical systems ($\{\Sigma_q\}_{q \in \mathbb{Q}}$) have a common $x_q$ and $u_q$.
        \end{assumption}

        \begin{assumption}
            At {every instant}, only an autonomous or a controlled jump can occur.
        \end{assumption}
        We discretize the differential equation governing ${x}_q$ and assume that in the small unit of time $dt$, $\dot{x}_q(x(t), u(t))$ remains constant. Also, the agent interacts with the environment after every $dt$ duration. Here, $dt$ is the design choice and can be chosen as small as possible. At a certain time $t$, state, action, model, reward and discount factor are given by:
     \newline
    \begin{itemize}[label = \(\circ\), leftmargin = *]
        \item State: ($x_t$, $q_t$)
        \item Action: ($u_t$, $v_t$, $x^1_t$, $q^1_t$)
        \item Model and Reward:
        \begin{algorithmic}
            \IF{$x_t \in B_{q_t}$}
                \STATE ($x_{t+1}, q_{t+1}) = G_{q_t}(x_t, q_t)$ \\
    			    $reward = R_{b}(x_t, q_t, v_t)$
            \ELSIF{$x_t \in C_{q_t}$}
                \STATE ($x_{t+1}, q_{t+1}$) = ($x^1_{t}, q^1_{t}$) \\
    			    $reward = R_{c}(x_t, q_t, x_{t+1}, q_{t+1})$
            \ELSE
                \STATE $q_{t+1} = q_t$ ; $x_{t+1}= x_t + \dot{x}_{q_{t+1}}(x_t, u_t) \times dt$   \\
    			   $reward = K(x_t, q_t, u_t) \times dt$
            \ENDIF
         \end{algorithmic}
        \item Discount Factor: $\gamma \in [0,1)$
    \end{itemize}
    \subsubsection{Examples}
    \label{ss:examples-type-I}
    \setcounter{ex}{0}
    \begin{ex}{(Revisited)}
    \label{ex:four-gear_car_extension}
    \rm{
        {We discuss two hybrid control design problems: Example \ref{ex:four-gear_car} and a similar scenario \textit{without} the impulse criteria (that resets the car velocity to $20$ m/s when it reaches $90$ m/s). We model both the settings in the proposed MDP framework.} We choose $dt$ as $0.01$. The agent needs to operate for $10$s, hence the total episode length is $1000$. As this is a finite horizon problem, the total episode reward is bounded, and we can choose the discount factor as $1$. At time $t$, state ($x_{1,t}, x_{2,t}$) corresponds to the position and velocity. Action ($u_t, v_t$) corresponds to the continuous and discrete controls. The model gives $q_{t+1}$, $x_{1,t+1}$ and $x_{2,t+1}$ as: $q_{t+1} = \min(\max(q_t + v_t, 1), 4)$. $x_{2,t+1} = x_{2,t} + f_{q_{t+1}}(x_{2,t}) u_t dt$ with an exception: for the problem with impulse, $x_{2,t+1} = 20$ when $x_{2,t} = 90$. Also, $x_{1,t+1} = x_{1,t} + x_{2,t+1} dt$. Reward is $x_{2,t} dt$. We execute PPO for $2000$ epochs for both problems. Figure \ref{CarGears_Results} discusses the results.\qed
    }
    \end{ex}
    \begin{figure*}[!htb]
    \centering
    \subfloat[Reward vs Epochs (\textit{Without} impulse)]{
    {\includegraphics[width=0.242\linewidth]{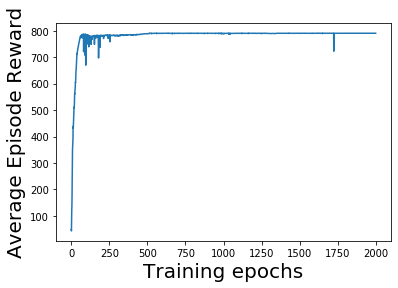}}
    }%
    \quad
    \subfloat[\textit{without} impulse: we plot car's gear position, throttle, and velocity/100 for $10$s using the policy at $3$ stages of training: at the start, after training for 200 epochs and at convergence (from left to right)]{
    {\includegraphics[width=0.249\linewidth]{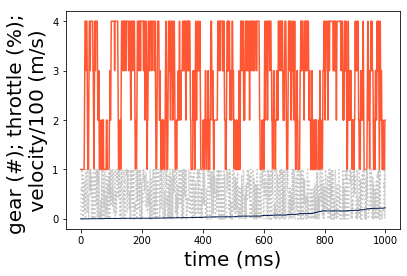}}
    {\includegraphics[width=0.22\linewidth]{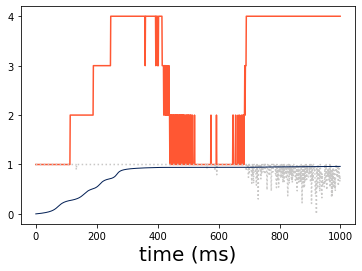}}
    {\includegraphics[width=0.22\linewidth]{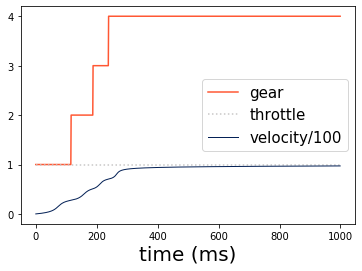}}
    }%

    \subfloat[Reward vs Epochs (\textit{With} impulse)]{
    {\includegraphics[width=0.242\linewidth]{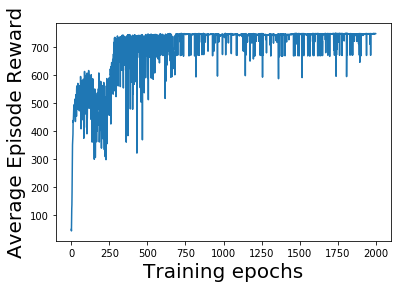}}
    }%
    \quad
    \subfloat[\textit{with} impulse: we plot throttle, car's gear position, and its velocity/100 for $10$s using the policy at $3$ stages of training: at the start, after training for 200 epochs and at convergence (from left to right)]{
    {\includegraphics[width=0.249\linewidth]{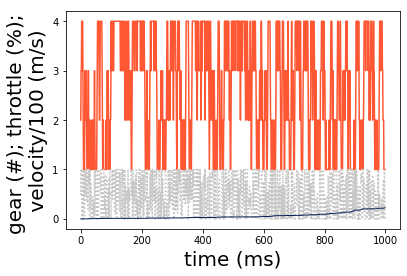}}
    {\includegraphics[width=0.22\linewidth]{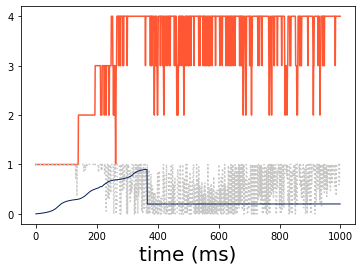}}
    {\includegraphics[width=0.22\linewidth]{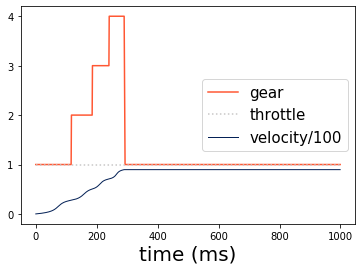}}
    }%
    \caption{(a) and (c) show that, for both problems, reward increases as the training progresses and eventually converges. (b) Plots for the problem without impulse show that, initially the agent changes gears and sets the throttle arbitrarily, thereby covering only a small distance. After training for $200$ epochs, the agent drives more sensibly. The policy at convergence shifts from gear $1$ to $2$ at the velocity at which $f_2(x_2)$ and $f_1(x_2)$ intersect. This makes sense as after this velocity, the second gear generates more acceleration than the first. Similarly, the agent changes from gear $2$ to $3$, and from gear $3$ to $4$, at velocities where $f_2(x_2)$, $f_3(x_2)$ and $f_3(x_2)$, $f_4(x_2)$ respectively intersect. (Refer to the figure \ref{CG_Velocity_Acceleration1} for $f_q(x_2)$). Also, the agent eventually learns to use full throttle during the entire episode. (d) Plots for the problem with impulse show that initially the agent drives randomly. After training for $200$ epochs, the agent drives more sensibly. However, the car reaches $90$ m/s, at which point its velocity is reset to $20$ m/s. The car gets stuck at this velocity and covers little distance in the remaining time. The last figure shows an amazing result. As per the policy at convergence, the agent \textsl{avoids} reaching $90$ m/s velocity. Instead, as soon as the car reaches $89.9$ m/s, the agent decreases the gear from $4^{\text{th}}$ to $1^{\text{st}}$ in its next $3$ subsequent interactions with the environment. As $f_1(89.9)$ is negligible, the car's velocity does not increase any more. The car cruises at this velocity for the remaining time, avoids reaching $90$ m/s, thereby tries to cover maximum distance.}
    \label{CarGears_Results}%
\end{figure*}

     \setcounter{ex}{2}
    \begin{ex}{(Water heater)}
    \label{ex:water-heater}
    \rm{
        {We consider the setting of a water heater from \cite{RaskinChapter}}. The example contains 3 components: (i) a tank that contains water, (ii) a gas burner that can be turned ON or OFF, and (iii) a thermometer that monitors the temperature of the water tank.

        Let {us} denote the temperature in the water tank by $x_1$, which is governed by: $\dot{x_1} = K(h-x_1)$ when the burner is ON and $\dot{x_1} = -Kx_1$ when the burner is OFF. We fix the value of $K$ to be $0.075$ and $h$ to be $150$. This rule is valid only when the temperature is within 20-100\textcelsius{}. Initial water temperature is $20$\textcelsius{}. Temperature cannot decrease beyond $20$\textcelsius{} and cannot increase beyond $100$\textcelsius{}.

        The goal of the agent is to operate for $30$s and stay within these three constraints: (C1) The temperature in the tank must never reach $100$\textcelsius{}. (C2) After $15$ seconds of operation, the system must be in stable regime, which is to say that the temperature in the water tank must always be between 91\textcelsius{} and 97\textcelsius{}. (C3) During the stable regime, the burner should never be continuously ON for more than $2$ seconds.

        We model this controlled switching problem in the MDP framework. We choose $dt$ as $0.01$. The agent has to operate for $30$s, hence the total episode length is $3000$. At time $t$, state ($x_{1,t}, x_{2,t}$) corresponds to the temperature of the water tank, and for how long the burner is \textit{continuously} ON. Action $v_t$ corresponds to the command to turn the burner ON or OFF, which takes a value $0$ or $1$. The model gives $x_{1,t+1}$ and $x_{2,t+1}$ as: $x_{1,t+1} = x_{1,t} + K(h-x_{1,t}) dt$; $x_{2,t+1} = x_{2,t} + dt$ when $v_t=1$, and $x_{1,t+1} = x_{1,t} -Kx_{1,t} dt$; $x_{2,t+1} = 0$ when $v_t = 0$. The reward is $R_1 + R_2 + R_3$. Here, $R_1 = -e^{(x_{1,t}-100)}$; $R_2 = \min(x_{1,t}-91, 97-x_{1,t},0)$ when $t \geq 15$s and $0$ otherwise; $R_3 = \min(2-x_{2,t}, 0)$ when $t \geq 15$s and $0$ otherwise. Here, we have defined our rewards in such a way that the agent obtains the maximum reward when the temperature profile satisfies all the conditions. We execute PPO for $3800$ epochs. Figure \ref{WaterHeater_Results} discusses the results.\qed
    }
    \end{ex}
    \begin{figure*}[!htb]
    \centering
    \subfloat[Reward vs Epochs]{
    {\includegraphics[width=0.245\linewidth]{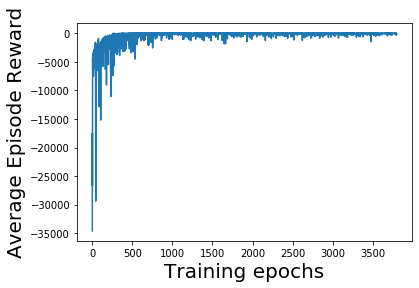}}
    }%
    \quad
    \subfloat[Temperature profile using the policy at $3$ stages of training: at the start, after training for 200 epochs and at convergence (from left to right)]{
    {\includegraphics[width=0.236\linewidth]{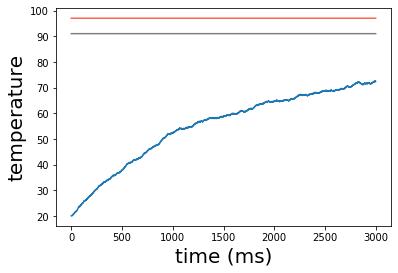}}
    {\includegraphics[width=0.226\linewidth]{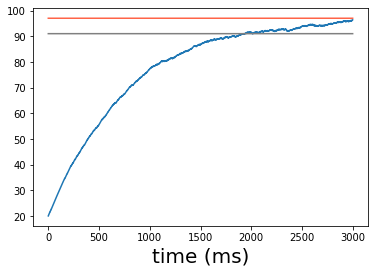}}
    {\includegraphics[width=0.226\linewidth]{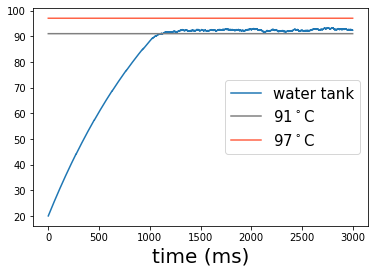}}
    }%
    \caption{(a) shows that the reward increases as the training progresses and eventually converges. (b) shows that, at the start of the training, the temperature profile violates several conditions. The profile gets better after training for $200$ epochs. The profile obtained by the policy at convergence does not violate any constraint, and the agent gets a reward close to $0$.}
    \label{WaterHeater_Results}%
    \end{figure*}

    {We now move on to our next set of results.}
\subsection{Type-\RNum{2} hybrid control design problems}
\label{ss:mainres2}
    Many manufacturing processes involve discrete entities (referred to as jobs) that pass through a series of workcenters. Physical state of these jobs (temperature, size, weight, chemical composition etc.) evolves according to certain time-driven dynamics, whereas temporal state of these jobs (waiting time or departure time of jobs at workcenters) changes according to some event-driven dynamics.
    Example \ref{ex:steel-annealing_process} is one such process. Quickly finishing processing a job may achieve the job completion deadline (good temporal state), but may not meet the target quality criteria (bad physical state). Achieving perfect quality (good physical state) may require more time for processing the job and runs the risk of not meeting deadlines (bad temporal state). Hence, a certain trade-off needs to be achieved.

    {In \cite{Cassandras1} the authors} propose a framework to model these problems as follows: The system starts at time $z_0$ from a physical state $\zeta_1$ and evolves according to the time-driven dynamics: $\dot{x_1} = f_1(x_1, u_1, t)$, where $u_1$ is a control. At time $z_1$, a switch takes place, which changes both the physical state as well as the dynamics. The physical state becomes $\zeta_2$ and the system evolves according to a new dynamics $\dot{x_2}$ from then on. In general, after the $i^{\text{th}}$ switch, the time-driven dynamics is given by $\dot{x_i} = f_i(x_i, u_i, t)$ with the starting condition $z_i(x_i) = \zeta_i$. The event-driven dynamics is given by $z_i = g_i(z_{i-1}, x_i, u_i, t)$. The optimisation problem in this case is the following:
    \begin{equation}
        \max_{u_1, .. ,u_N} \sum_{i=1}^{N} L_i(z_i, u_i).
        \label{Eq_Cassandras_Equations}
    \end{equation}
    Here, $N$ is the total number of jobs. Notice that (\ref{Eq_SA_non_linear_heat_transfer1}), (\ref{Eq_SA_Temporal_State1}) and (\ref{Eq_SA_Optimisation_Equation1}) from Example \ref{ex:steel-annealing_process} represent time-driven dynamics, event-driven dynamics and optimisation problem respectively. For simplicity, the framework assumes that the control variables are time independent. Hence, the agent gets to observe only the arrival times $a_i$ of jobs, and has to decide $u_i$ straightaway.
    \subsubsection{MDP Framework}
    \label{ss:mdp-type-II}
        We model {Type-II hybrid control design} problems as Markov Decision Processes with the state being ($a_1, a_2, \ldots, a_N$), action being ($u_1, u_2, \ldots, u_N$) and the reward being $\sum\limits_{i=1}^{N} L_i(z_i, u_i)$. The key here is the assumption that the control variables are time-independent, which makes it a single-stage process. Thus, there is no next state, and hence no model. Notice that $u$ in equation (\ref{Eq_SA_non_linear_heat_transfer1}) is time-independent. Hence, once the controller sets the velocities $u_i$, it cannot interfere. Also, as the length of the episode is $1$, the discount factor is also redundant. Hence, this is a contextual bandit problem \citep{SzepLattimore_book}.
    \subsubsection{Examples}
    \label{ss:examples-type-II}
    \setcounter{ex}{1}
    \begin{ex}{(Revisited)}
    \label{ex:single-furnace}
    \rm{
        Example \ref{ex:steel-annealing_process} is a steel annealing process with a single furnace. We solve the same problem mentioned in \cite{SteelAnnealing}. Five strips arrive at $0.2$, $2$, $4$, $7$ and $10$ minutes respectively, with all of them having length of $1000$m. $u_{\min} = 100$ m/min and $u_{\max} = 700$ m/min. We modelled this problem in the MDP framework and ran the PPO algorithm for $5000$ epochs. Figure \ref{SASF_Results} discusses these results.\qed
    }
    \end{ex}
   \begin{figure*}[!htb]
    \centering
    \subfloat[Reward vs Epochs]{
    {\includegraphics[width=0.245\linewidth]{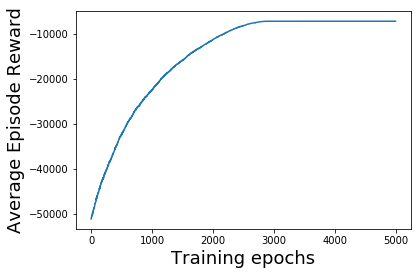}}
    }%
    \quad
    \subfloat[Temperature profile for policies at 3 stages of training: at the start, after training for 200 epochs and at convergence (from left to right)]{
    {\includegraphics[width=0.236\linewidth]{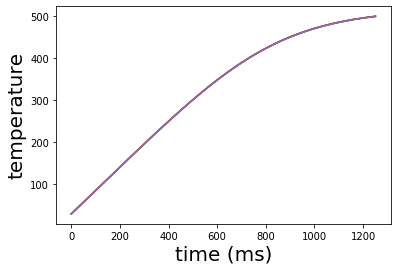}}
    {\includegraphics[width=0.226\linewidth]{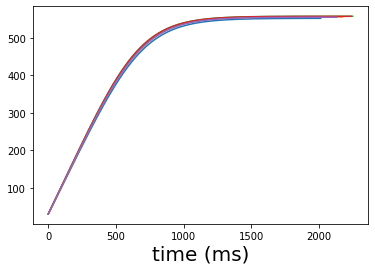}}
    {\includegraphics[width=0.226\linewidth]{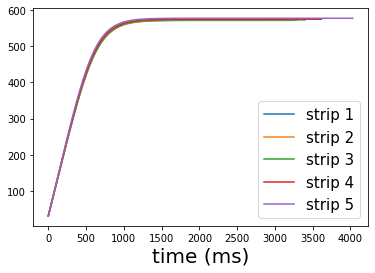}}
    }%
    \caption{(a) shows that the reward increases as training progresses and eventually converges. (b) shows that initially the strips are not able to reach the furnace reference temperature of 600\textcelsius{}. The temperature profile gets better after $200$ epochs. Finally, all strips are able to reach the reference temperature.}
    \label{SASF_Results}%
    \end{figure*}
    \setcounter{ex}{3}
    \begin{ex}{(Multiple furnace systems)}
    \label{ex:multiple-furnace}
    \rm{
        Some annealing processes have multiple furnaces, such as a heating furnace, a soaking furnace and a cooling furnace, operating in tandem. These furnaces may have different reference temperatures and different furnace lengths. The thermal dynamics (physical states) in each furnace is governed by \eqref{Eq_SA_non_linear_heat_transfer1} with corresponding initial states.

        {We} consider $N$ strips and $M$ furnaces with a total of $N \times M$ jobs. Let $h_i$ and $a_i$ be the length and arrival time of the $ith$ strip respectively, $L_j$ be the length of $jth$ furnace and $u_{i,j}$ be the velocity of $ith$ strip in the $jth$ furnace. Let $z_{1,0,1} = -\infty$. The temporal states $z_i$ are computed by
         \begin{equation}
        \begin{aligned}
            z_{1,i,1} = \max(a_i, z_{1,i-1,1}) +\frac{h_i}{u_{i,1}}, \\
            z_{2,i,3} =  z_{1,i,1} + \frac{L_1}{u_{i,1}} +  \sum_{k=2}^{M}\frac{h_i + L_k}{u_{i,k}},  \\
        \end{aligned}
        \label{Eq_SAMF_Temporal_State}
        \end{equation}
        where $z_{1,i,1}$ denotes the time at which the $ith$ strip completely enters the first furnace, and $z_{2,i,3}$ denotes the time at which it leaves from the third furnace.

        The optimal control problem can be described by
    \begin{equation}
        \begin{aligned}
            \max_{u_{1,1}, .. ,u_{N,M}} \sum_{i=1}^{N}&\:\:[ \:\phi(z_i) + (\sum_{j=1}^{M}{\theta(u_{i,j}))}] \\
            \sbjto\:\:&\:\:(\ref{Eq_SA_approx_heat_transfer1}), (\ref{Eq_SAMF_Temporal_State}), (\ref{Eq_SAMF_Costs}),
        \end{aligned}
        \label{Eq_SAMF_Optimisation_Equation}
    \end{equation}
    where
    \begin{equation}
        \begin{aligned}
            \phi(z_i)  &=  -(z_{2,i,3})^2, \\
            \theta(u_{i,j}) &=  -\abs{F_j - x_i(L_j/u_{i,j})}^2 - \beta \int_0^{L_j/u_{i,j}}(F_j - x_i(t))^2 \;dt.
        \end{aligned}
        \label{Eq_SAMF_Costs}
    \end{equation}

    We solve the same problem mentioned in \cite{SteelAnnealing}. Six strips arrive at $0.8$, $8$, $12$, $16$, $24$ and $28$ minutes respectively, with all of them having length of $600$m. $3$ furnaces have length $400$, $300$, $200$m and reference temperatures $500$, $400$, $200$\textcelsius{} respectively. $u_{\min} = 100$ m/min and $u_{\max} = 700$ m/min. We modelled this problem in the MDP framework and ran the PPO algorithm for $20000$ epochs. Figure \ref{SAMF_Results} discusses the results.\qed
    }
    \end{ex}

\begin{figure*}[!htb]
    \centering
    \subfloat[Reward vs Epochs]{
    {\includegraphics[width=0.245\linewidth]{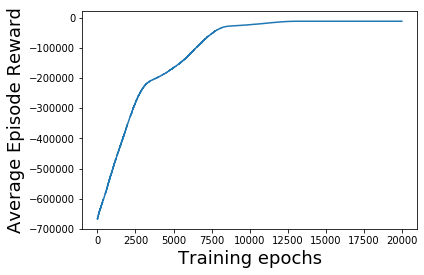}}
    }%
    \quad
    \subfloat[Temperature profile for policies at 3 stages of training: at the start, after training for 5000 epochs and at convergence (from left to right)]{
    {\includegraphics[width=0.236\linewidth]{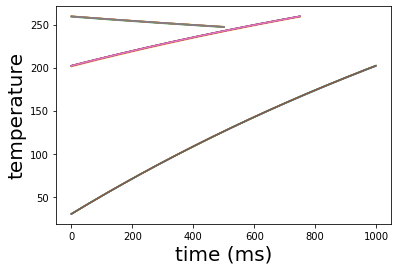}}
    {\includegraphics[width=0.226\linewidth]{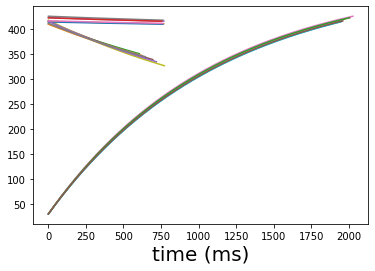}}
    {\includegraphics[width=0.226\linewidth]{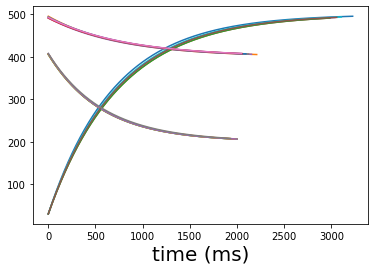}}
    }%
    \caption{(a) shows that the reward increases as training progresses and eventually converges. (b) shows that initially no strip is able to reach the reference temperature in any furnace. Eventually, all strips reach the reference temperature in all furnaces.}
    \label{SAMF_Results}%
\end{figure*}
\subsection{Discussion}
\label{ss:discssn}
    So far, we have presented MDP frameworks for both Type-I and Type-II hybrid control design problems, modelled a set of benchmark examples in the proposed frameworks, and designed optimal control policies by employing PPO algorithm from RL literature. We now highlight a few features of our results.
    \begin{rem}
    \label{rem:no_model}
    \rm{
        Hybrid dynamical systems cover a very large class of practical settings under its canopy. Not surprisingly, researchers in systems and control domain have proposed various frameworks to model a hybrid system for control design, see e.g., \cite{Bemporad, BBBM, Branicky1, Cassandras1, Hedlund, SupervisoryKA, Piccoli, Sussmann, Tomlin, Koutsoukos2005, Platzer2011} for detailed discussions. In this paper we restrict our attention to the frameworks from \cite{Branicky1} and \cite{Cassandras1} (described in \S\ref{ss:mainres1} and \S\ref{ss:mainres2}, respectively) to develop our results.
    }
    \end{rem}
    \begin{rem}
    \label{rem:large_class}
    \rm{
        The proposed techniques on model-free hybrid control design cater to a large class of hybrid systems. Indeed, we consider various classes of ``jumps'' in the dynamics (including impulses and switches, both for autonomous and controlled cases) and do not opt for continuous or discrete abstractions of the hybrid action space in our design of optimal control policies.
    }
    \end{rem}
\section{Conclusion}
\label{s:concln}
    {In this paper we presented a model-free approach to the design of hybrid control policies for two classes of hybrid systems using RL techniques. In particular, for each case, we proposed MDP frameworks, modelled a set of benchmark examples in that framework, and applied the PPO algorithm to design optimal control policies. It is well-known that a large class of modern day Cyber-Physical Systems (CPS) can be modelled as hybrid dynamical systems. A next natural topic of investigation is to study the performance of our framework in complex large-scale CPS settings. This matter is currently under investigation, and will be reported elsewhere.}
\bibliographystyle{plain}
\bibliography{HybridControl_References}
\end{document}